\documentclass[%
 reprint,
 amsmath,amssymb,
 aps,
]{revtex4-1}

\usepackage{graphicx}
\usepackage{dcolumn}
\usepackage{bm}
\usepackage{braket}
\usepackage{here}
\usepackage{wrapfig}
\usepackage{color}
\usepackage{comment}

\begin{document}

\preprint{APS/123-QED}

\title{Propagation of phase-imprinted solitons from superfluid core \\ to Mott-insulator shell and superfluid shell}

\author{Yuma Watanabe, Shohei Watabe and Tetsuro Nikuni}%
\affiliation{
Department of Physics, Faculty of Science Division I, Tokyo University of Science, Shinjuku, Tokyo,162-8601, Japan
}

\begin{abstract}
	We study phase-imprinted solitons of ultracold bosons in an optical lattice with a harmonic trap exhibiting superfluid (SF) and Mott-insulator (MI) shell structures. 
	An earlier study [Konstantin V. Krutitsky, J. Larson, and M. Lewenstein, Phys. Rev. A 82, 033618 (2010).] reported three types of phase-imprinted solitons in the Bose--Hubbard model: in-phase soliton, out-of-phase soliton, and wavelet. 
	In this paper, we uncover the dynamical phase diagram of these phase-imprinted solitons. 
	We also reveal another type of phase-imprinted soliton, the hybrid soliton. 
	In a harmonically trapped system, solitonic excitations created at the SF core cannot penetrate the outer SF shell. 
	This repulsion at the surface of the outer SF shell can be counteracted by imposing a repulsive potential at the center of the trap. 
	These results can be interpreted as a kind of impedance matching of excitations in Bose--Einstein condensates in terms of the effective chemical potentials or local particle numbers in the shell, and analogous results can be observed in sound waves created by a local single-shot pulse potential. 
\end{abstract} 
\maketitle


\section{\label{sec:level1}INTRODUCTION}
Ultracold atomic gases provide an ideal platform for studying many-body phenomena, where quantum fluctuations play a crucial role.
	In particular, optical lattices enable us to realize the Bose--Hubbard (BH) model, which exhibits a fundamental quantum phase transition known as the superfluid (SF)--Mott-insulator (MI) transition~\cite{Jaksch1998, Greiner2002}. 
	In the SF state, each atom is delocalized over all lattice sites and a many-body state is well described by a macroscopic wave function with long-range phase coherence. 
	In contrast, individual atoms are localized at each site in the MI state, due to the strong repulsive interaction, compared with the hopping of atoms~\cite{Fisher1989}.
	In the presence of a harmonic trap, these states can coexist, forming a shell-structure, where the SF and MI regions emerge alternately~\cite{Batrouni1990, Batrouni2002, Foelling2006}.
	The controllability of the system parameters is an advantage of ultracold atoms that makes the system a perfect playground for studying the matter--wave nature as well as nonlinear phenomena, such as solitons~\cite{Pitaevskii2003, Kivshar2003}.
	
	Solitons are solitary waves that are robust and persistent and emerge in many aspects of nature.
	In the case of weakly interacting Bose condensates in one dimension, the Gross--Pitaevskii (GP) equation~\cite{Gross1961, Pitaevskii1961,Ablowitz1991} describing the system has a stable soliton solution.
	In the case of an attractive interaction, the GP equation has bright-soliton solutions~\cite{Khaykovich2002, Strecker2002}, while in the repulsive case the dark-soliton solution exists~\cite{Pethick2008, Pitaevskii2003}.
	In atomic superfluids, dark solitons are characterized by a localized density dip and a phase kink in the complex order parameter.
	Dark solitons have been experimentally observed in Bose--Einstein condensates (BECs) with the phase-imprinting method~\cite{Denschlag2000, Burger1999, Stellmer2008}, density-engineering method~\cite{Burger2002}, and merging of two coherent condensates~\cite{Weller2008}.
	Bright and dark solitons in one-dimensional Bose gases have been theoretically investigated widely and intensively ~\cite{Delande2014, Mishmash2009, Hebert2005, Dziarmaga2004, Delande2014, Shamailov2019, Krutitsky2010}.

The features of stationary standing solitons have been studied in homogeneous lattice systems near MI lobes~\cite{Krutitsky2010}. 
Three types of stationary standing soliton, characterized by a peak, a dip, and a dip pair bonding with respect to the mean particle occupation number, can be created in the SF state. 
The phase imprinting method is useful for creating solitons and has been implemented in experiments in ultracold atoms~\cite{Burger1999}. 
This method can create a dynamical excitation (not a stationary standing soliton), where a peak and dip with respect to the mean particle occupation number move in opposite directions~\cite{Krutitsky2010}. 
An earlier study~\cite{Krutitsky2010} reported three different types of phase imprinted solitons, which can be classified by the structures of the local condensate particle number and the local total particle number.
The creation of these three types of excitations in the condensate depends on system parameters such as the chemical potential, interaction strength, and hopping parameter.

There remain several unsolved questions regarding the properties of solitons in optical lattices.
First, in uniform systems, the three types of solitons have been found using only a few combinations of system parameters~\cite{Krutitsky2010}. 
The classes of phase imprinted solitons have therefore not been exhaustively explored and a phase diagram for the structure of solitons has not been determined.
Second, controlling the structure of phase-imprinted solitons and transport excitations in an inhomogeneous background is not well understood but will be important in the context of recent developments in atomtronics~\cite{amico2020roadmap}. 
Since the local chemical potential effectively changes in a harmonically trapped system, an experimental setup may provide the SF--MI--SF shell structure, which enables us to study excitations propagating in an SF--MI--SF heterojunction.

	In this paper, we study the dynamics of phase-imprinted solitons in the Bose--Hubbard model using the time-dependent Gutzwiller approximation.
	We first investigate the structure of phase-imprinted solitons in the uniform Bose--Hubbard model and determine the dynamical phase diagram. 
	In addition to the three types of phase imprinted solitons~\cite{Krutitsky2010}, we find a new type of phase imprinted soliton that was not reported in the earlier paper~\cite{Krutitsky2010}. 
	We also investigate the dynamics of phase imprinted solitons in an SF--MI--SF heterojunction. 
	We create phase-imprinted solitons in the SF core at the center of a harmonic trap and trace the subsequent dynamics of the solitonic excitations, which propagate toward the outer MI and SF shells.
	We find that the excitations propagate and permeate the MI regions. 
	However, these excitations cannot percolate into the outer SF shell and are reflected by its surface. 
	In the presence of an additional repulsive potential at the center of the trap, creating a double-well trap, 
	we can realize an SF--MI--SF heterojunction where the same type of solitons can propagate in the SF core and outer SF shell.
	This is the only case that we found in which the excitation created in the SF core can percolate into the outer SF shell.

	
		\begin{figure*}[tb]
		\centering
			\includegraphics[scale=0.3]{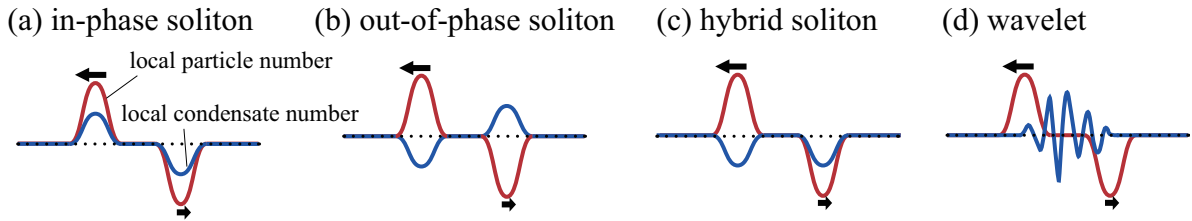}
			\caption{
			Schematics of phase imprinted solitons:
			(a) in-phase soliton, (b) out-of-phase soliton, (c) hybrid soliton, and (d) wavelet.
			The red and blue lines represent the structures of the local particle number and local condensate number in these solitons, respectively.
			The propagation speeds of the dip and peak structures are depicted by the length of the arrows, where the speed of the peak structure is faster than that of the dip structure.
			We note that although the local particle and condensate numbers are functions of the discrete site index $l$, we depict them as continuous quantities for clarity.
			}	
			\label{figure_0_1}
		\end{figure*}

\section{\label{sec:level1}model}
	The system of an ultracold Bose gas in an optical lattice is well described by the Bose--Hubbard (BH) Hamiltonian~\cite{Fisher1989, Greiner2002}:
		\begin{equation}
		\hat{H}=-J\sum_{\braket{l, m}}\hat{b}^\dagger_l\hat{b}_m+\sum_l(\epsilon_l-\mu)\hat{n}_l+\frac{U}{2}\sum_l\hat{n}_l(\hat{n}_l-1), 
		\end{equation}
	where $U$ is the atom--atom interaction strength, $J$ is the hopping coefficient, $\mu$ is the chemical potential, $\epsilon_l$ is the on-site energy that includes the effect of the external potential, $\hat{n}_l\equiv\hat{b}_l^\dagger\hat{b}_l$ is the number operator at site $l$, and $\hat{b}_l$ and $\hat{b}_l^\dagger$ are annihilation and creation operators that obey the bosonic commutation relation $\left[\hat{b}_l, \hat{b}_m^\dagger\right]=\delta_{l, m}$. 
	We consider a one-dimensional optical lattice system containing $L$ sites.


	To investigate the time evolution in the one-dimensional BH model, we use the Gutzwiller approximation.
	In the Gutzwiller ansatz, the state $\ket{\Phi}$ given by the BH model is represented as 		
		\begin{equation}
		\ket{\Phi}=\prod_l\sum_{n=0}^{\infty} f_{n}^{(l)}\ket{n}_l, 
		\end{equation}
	where $\ket{n}_l$ is the local Fock state.
	The equation of motion for the coefficient $f_n^{(l)}(t)$ can be derived from the stationary condition of the effective action
		\begin{equation}
		S=\int dt\Braket{\Phi|i\hbar\frac{d}{dt}-\hat{H}|\Phi} . 
		\label{eq4}
		\end{equation}	
	
		Using the Gutzwiller variational state $\Ket{\Phi}$ in Eq.~(\ref{eq4}) and imposing the stationary condition on $S$ with respect to the variation of the Gutzwiller coefficient $f_n^{(i)}$, we find
		\begin{align}
		i\hbar\frac{df^{(l)}_n}{dt}=&-J\sum_{\braket{m}_l}\left(\Phi_m\sqrt{n}f_{n-1}^{(l)}+\Phi^\ast_m\sqrt{n+1}f_{n+1}^{(l)}\right)
		\nonumber 
		\\
		& +\left[\frac{U}{2}n(n-1)-\mu_l n\right]f_n^{(l)} , 
		\label{eq:EOM_GutzwillerCoefficient}
		\end{align} 
		where $\langle m\rangle_l$ denotes the nearest-neighbor sites for the site $l$.
		 Here, $\Phi_l\equiv\braket{\Phi|\hat{b}_l|\Phi}$ is the superfluid order parameter.
		 Within the Gutzwiller approximation described by Eq.~(2), $\Phi_l$ is given by
		 \begin{align}
		 \Phi_l=\sum_{n=0}^{\infty} f_{n}^{(l)\ast}\sqrt{n+1}f_{n+1}^{(l)} .
		 \label{eq:OrderParameter}
		 \end{align}
		 We define the effective chemical potential as $\mu_l=\mu -\epsilon_l$.
	To investigate the characteristic features of the solitons, we calculate the evolution of the local particle number
		\begin{align}
		n_l\equiv\braket{\Phi|\hat{n}_l|\Phi}=\sum_{n=0}^\infty n|f_n^{(l)}|^2 ,
		\end{align} 
	and local condensate number $n_{cl}=|\Phi_l|^2$.
	The non-local condensate number is given by   $\tilde{n}_l = n_l - n_{cl}$.
	
	Strictly speaking, the Gutzwiller approximation fails in one dimension except for very weak interactions $U/J \ll 1$.
	Nevertheless, this approximation has been widely used to study the dynamical properties of the one-dimensional Bose--Hubbard model \cite{Altman2005, Polkovnikov2005, Saito2012}, since it is very useful for illustrating the basic physics.
	We also note that Ref.~\cite{Krutitsky2010} starts with the three-dimensional Bose--Hubbard model and assumes that the Gutzwiller coefficient $f_n$ and order parameter $\Phi$ depend only on one spatial dimension.
	This assumption leads to an equation of motion that is essentially the same as Eq.~(\ref{eq:EOM_GutzwillerCoefficient}) (see Appendix A).
	Therefore, our analysis using the Gutzwiller approximation will be useful in determining the dynamical properties of the three-dimensional model, which is more realistic in describing the experimental situation.	
	
	Dark and grey solitons are experimentally created by the phase-imprinting method~\cite{Denschlag2000, Burger1999}. 
	In this method, we apply a laser beam to half of the system for a short time $t_{\mathrm{imp}}$, which induces a spatial shift in the phase of the order parameter, creating the solitons. 
	In the BH model, the effect of the phase-imprinting laser beam can be included in the on-site energy term  $\sum_l\epsilon_l\hat{n}_l$. 
	Within the Gutzwiller approximation, if the imprinting potential is applied for a very short time compared to the other characteristic time scales, we can simply include the effect of the imprinting potential using
	\begin{equation}
	f_{n}^{(l)}(t_{\mathrm{imp}})=f_{n}^{(l)}(t=0)e^{-i\phi_l n} , \ \ \ \phi_l=\frac{\epsilon_l t_\mathrm{imp}}{\hbar} .
	\label{eq6}
	\end{equation}
	We choose a hyperbolic tangent imprinting potential as used in Ref.~\cite{Krutitsky2010}:
	\begin{equation}
	\phi_l=\frac{\epsilon_l t_{\mathrm{imp}}}{\hbar}=\frac{\Delta\phi}{2}\left[1+\mathrm{tanh}\left(\frac{l-l_0}{0.45l_{\mathrm{imp}}}\right)\right].
	\label{eq7}
	\end{equation}
	Here, $l_0$ is the center of the system, $l_{\mathrm{imp}}$ is the width characterizing the potential variation, and $\Delta\phi$ represents the amplitude of the imprinting potential~\cite{Krutitsky2010}.
	In our simulation, we first obtain the ground state
	and then prepare the initial phase-imprinted state using the Gutzwiller coefficients with Eqs.~\eqref{eq6} and~\eqref{eq7}.


\section{\label{sec:PD}dynamical phase diagram}
		
		\begin{figure}[bt]
		\centering
			\includegraphics[scale=0.25]{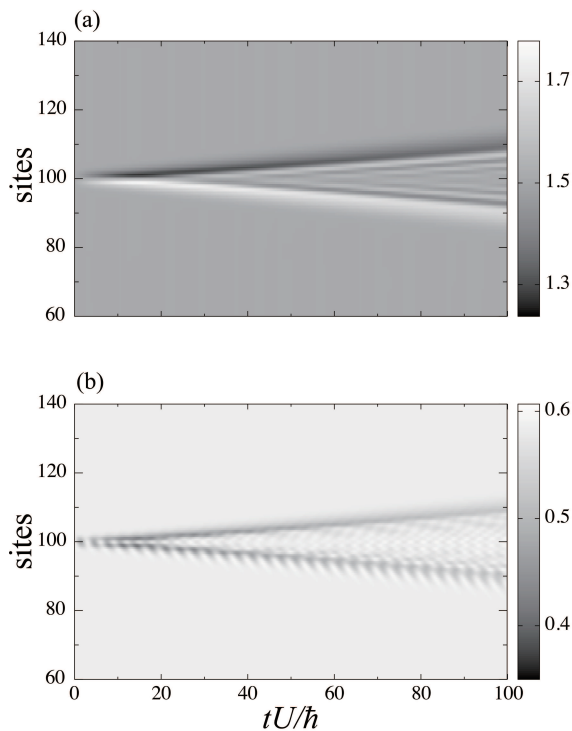}
			\caption{
			Time evolution of the local particle number (a) and the local condensate number (b) characteristic of the hybrid soliton.
			We used $J/U=0.020$ and $\mu/U=1.0$.
			The local particle number has a dip and a peak which propagate in opposite directions.
			The local condensate number, on the other hand, has two dips, which propagate in opposite directions. 
			}
			\label{figure_0_2}
		\end{figure} 
		
		\begin{figure}[bt]
		\centering
			\includegraphics[scale=0.925]{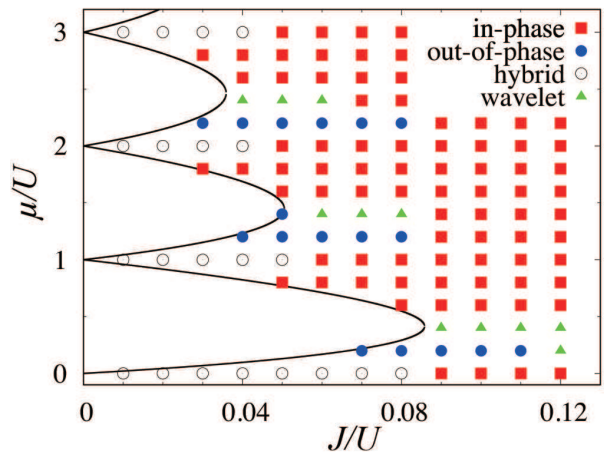}
			\caption{
			Dynamical phase diagram of the phase-imprinted solitons: in-phase solitons (red squares), out-of-phase soliton (blue circles), hybrid solitons (white circles), and wavelets (green triangles). 
			}
			\label{figure_0_3}
		\end{figure}
		
	In this section, we discuss the dynamical phase diagram of phase-imprinted solitons in the uniform BH model.
	In Ref.~\cite{Krutitsky2010}, two types of modes, the on-site mode and off-site mode, are reported for the stationary soliton in an optical lattice system~\cite{Krutitsky2010}.
	In the on-site mode, a phase kink is placed at the lattice site.
	In the off-site mode, a phase kink is placed between two nearest sites.
	In this paper, we focus on the on-site mode soliton because the lifetime of on-site mode solitons is longer than that of off-site mode solitons~\cite{Krutitsky2010}. 
	
	As for the dynamical properties, three types of soliton dynamics have been reported in one-dimensional uniform systems~\cite{Krutitsky2010}.
	However, they have been studied using only a few combinations of system parameters. 
	We here exhaustively investigate the class of soliton dynamics by systematically changing the system parameters in the $J/U$--$\mu/U$ plane 
	and determine the dynamical phase diagram of phase-imprinted solitons in the uniform Bose--Hubbard model. 	
	We simulate the soliton dynamics created by the phase-imprinting method in a one-dimensional uniform system within the time-dependent Gutzwiller approximation.
	We calculate the propagation of the local particle number and local condensate number, 
	which enables us to classify the dynamics of the solitons.
	
	We find four types of dynamics, as shown in Fig.~\ref{figure_0_1}: 
	(a) in-phase soliton, (b) out-of-phase soliton, (c) hybrid soliton, and (d) wavelet.
	Of these, types (a), (b), and (d) have already been reported in Ref.~\cite{Krutitsky2010}, 
	 while (c), which we call the hybrid soliton, is a new type of soliton dynamics.

	In all types of soliton dynamics, the local particle number has a peak and a dip, which propagate in opposite directions. 
	The peak propagates faster than the dip.
	The four types of soliton dynamics are characterized by the structure of the local condensate number. 
	In the in-phase soliton (Fig.~\ref{figure_0_1} (a)), the condensate peak and dip are situated at the peak and dip in the local particle number. 
	In the out-of-phase soliton (Fig.~\ref{figure_0_1} (b)), the condensate peak and dip emerge at the dip and peak, respectively, of the local particle number. 
	In the wavelet (Fig.~\ref{figure_0_1} (d)), although the local particle number exhibits a peak and dip structure, we cannot find any peak or dip solitary structure in the local condensate number, where the small wavelet oscillation emerges. 
	
	The hybrid soliton (Fig.~\ref{figure_0_1} (c)) exhibits two dips in the local condensate number, which emerge and propagate in opposite directions, while the local particle number has a peak and dip. 
	We show the dynamics of the local particle number and the local condensate number in Fig.~\ref{figure_0_2} with $J/U=0.020$ and $\mu/U=1.0$. 
		
	Figure~\ref{figure_0_3} shows a dynamical phase diagram for the four types of phase-imprinted soliton.
	The in-phase soliton region is widely distributed in the phase diagram. 
	The out-of-phase soliton, hybrid soliton, and wavelet emerge close to the MI lobes, where $J/U$ is small. 
	The hybrid soliton region encompasses areas where the chemical potential $\mu/U$ is close to an integer value.
	The wavelet (d) is situated  in the narrow regions between the in-phase soliton and out-of-phase soliton regions. 
	
	The out-of-phase solitons can be interpreted as dark solitons of holes~\cite{Krutitsky2010}.
	This type of soliton appears in the ``hole area" in the phase diagram (see Fig.~2 of Ref.~\cite{Krutitsky2010}), where the superfluidity has a hole character.
	We note that the hybrid soliton is situated in the narrow region near the phase boundary between the hole area and particle area.
	We can interpret the hybrid soliton as simultaneously excited particle and hole solitons propagating in opposite directions.
	
	We also note that in obtaining the dynamical phase diagram, we identified different types of solitons by examining the spatial profiles of particle number and condensate number.
	However, there are no apparent criteria that can determine the phase boundaries analytically.
	In fact, two different types of soliton dynamics (for example, in-phase soliton and wavelet) emerge simultaneously near the phase boundary.
	Similar phenomena, i.e., the emergence of two different dynamics, have also been found in Ref.~\cite{Krutitsky2010}.
	In the region where the in-phase or out-of-phase soliton and the wavelet coexist, we can distinguish between the in-phase or out-of-phase soliton and the wavelet by comparing their speeds.
	When the propagation speeds of the peak and dip are faster than the spreading speed of the wavelet, 
	we identified this as the in-phase or out-of-phase soliton.
	However, near the phase boundary, the two speeds are comparable and it is hard to make a clear distinction.
	It is therefore necessary to employ other approaches to determine a more precise phase boundary in the dynamical phase diagram. 

	Finally, we discuss the possibility of the experimental observation of the four types of solitons.
	As proposed in Ref.~\cite{Krutitsky2010}, the four types of solitons discussed in this section could be observed via the time-of-flight method~\cite{Bloch2008}, 
	which measures the momentum distribution $P(k)$.
	Within the Gutzwiller approximation, the momentum distribution is given by~\cite{Krutitsky2010}
		\begin{align}
		P(k)=|W(k)|^2\left[\sum_l(\braket{n}_l-|\Phi_l|^2)+\left|\sum_l\Phi_le^{-ikl}\right|^2\right] ,
		\end{align}
	where $W(k)$ is the Fourier transform of the Wannier function.
	Within the mean-field approximation, the condensates component dominantly contributes to the momentum distribution.
	Therefore, we may expect that the characteristic features of solitons show up in the momentum distribution reflecting the characteristic features in the condensate order parameter $\Phi_l$.

\section{\label{sec:level2}Excitation transport from SF core to MI and SF shells}

We study the transport of the solitonic excitation in a heterojunction of the SF--MI--SF state in this section. 
In particular, a Bose gas in an optical lattice confined in a harmonic trap provides the shell structure of the SF and MI states. 
We create a phase-imprinted soliton at the SF core and study the dynamics of an excitation propagating to the outer MI shell and outer SF shell. 
	We consider a one-dimensional optical lattice system containing $L$ sites combined with a harmonic trap potential, and we choose the on-site energy $\epsilon_l$ to be 
		\begin{equation}
		\epsilon_l=V_0 \left(l-l_0\right)^2,
		\end{equation}
	where $V_0$ is the curvature of the confining harmonic potential.
	In the presence of a harmonic trap, the system exhibits a shell-structure of the SF and MI regions, which are placed alternately. 
	
	We calculate the dynamics of the solitons created by the phase-imprinting method within the time-dependent Gutzwiller approximation.
	Initially, we calculate the ground state by an imaginary time relaxation in a presence of a harmonic trap.
	We then apply the imprinting potential using Eqs.~(\ref{eq6}) and (\ref{eq7}), which creates a soliton at the center of the SF core. 
	To investigate the dynamics of the solitons, we calculate the propagation of the local particle number and the condensate.
	 We use $V_0/ U= 5.0\times 10^{-4}$ and $J/U=0.070$. 
	The chemical potential is chosen to be $\mu/U=0.75$ for the in-phase soliton, $\mu/U=1.2$ for the out-of-phase soliton, $\mu/U=1.0$ for the hybrid soliton, and $\mu/U=1.4$ for the wavelet (see Fig.~\ref{figure_0_1}).
	The local condensate order parameter $\Phi_l$ has a very small value at all the sites in the Mott region within our Gutzwiller approximation. 
	In the present paper, we regard the MI region as $n_{cl}<0.05$, where $n_{cl}=|\Phi_l|^2$ is the local condensate particle number.
	
	We simulate the propagation of all the types of solitons shown in Fig.~\ref{figure_0_1} using the parameters given above.
	In the SF core, we can observe the characteristic structures of the solitons at the beginning of propagation. 
	However, we find that the results of the transport of the out-of-phase soliton, hybrid soliton, and wavelet to the outer MI and SF shells are qualitatively the same as that of the in-phase soliton. 
	This is due to the effect of a harmonic trap. 
	The effective chemical potential monotonically decreases from the center of the trap to the outer region, 
	and the in-phase soliton regions emerges in the SF phase at the upper side of the MI region, as shown in Fig.~\ref{figure_0_3}. 
        As a result, we find that even if we create an out-of-phase soliton, hybrid soliton, and wavelet, we end up injecting an in-phase soliton to the MI region, because all the types of phase-excitations turn into an in-phase soliton before the excitation reaches the outer MI shell. 
	In the following, we thus show results of an in-phase soliton created in the SF core. 
	
	\begin{figure}[tb]
	\centering
		\includegraphics[scale=0.7]{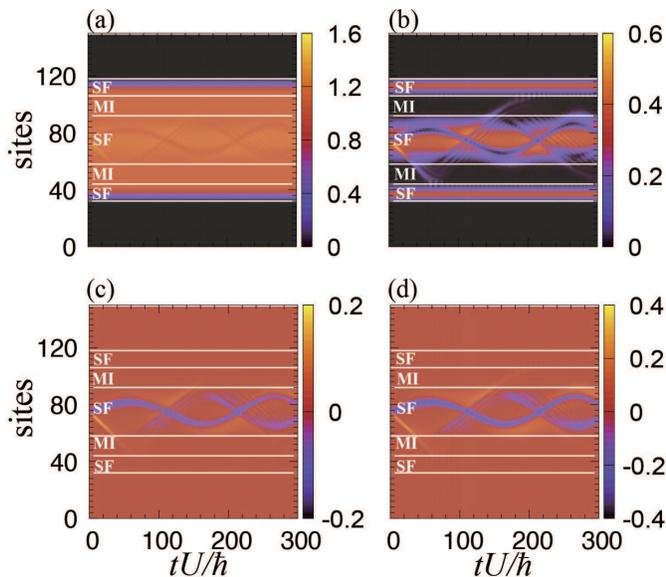}
		\caption{
		 Time evolution of the local particle number (a) and local condensate particle number (b). 
		 The differences of the local particle number (c) and local condensate particle number (d) from the initial state. 
		 We used $J/U=0.070$, $\mu/U=0.750$, and $V_0/ U= 5.0\times 10^{-4}$. 
		}
		\label{figure_1_1}	
	\end{figure}
		
		In Fig.~\ref{figure_1_1}, we plot the time evolution of (a) the local particle number $n_l$, (b) local condensate particle number $n_{cl}$, (c) difference of the local particle number from the initial state, and (d) that of the local condensate particle number. 
	In the SF core region, the positions of the dips and peaks of the local condensate particle number are the same as those of the local particle number (see Fig.~\ref{figure_1_1}). 
	The solitons collide with the surface of the MI shell and the condensates propagate in the MI shell.
	The excitations then reach the boundary between the MI shell and outer SF shell.
	The excitations do not penetrate into the outer SF region and are reflected by its surface. 
	Then, the oscillation at the surface of the SF shell emerges. 
	This can be seen more clearly in the case where the peak (not the dip) of the soliton collides with the outer MI shell. 
	
\begin{figure}[htb]
	\centering
		\includegraphics[scale=0.5]{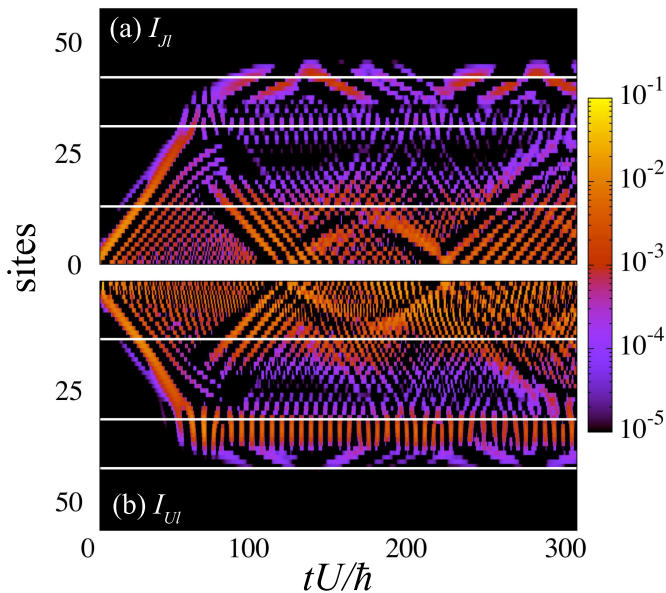}
		\caption
		{
		Comparison of the contributions from $I_{Jl}$ (a) and $I_{Ul}$ (b) in Eq.~(\ref{eq_CC}).
		The parameters are the same as in Fig.~\ref{figure_1_1}.
		}
		\label{figure_0_5}
	\end{figure}

	To understand the dynamics of the condensate in more detail, we derive the equation of motion for the condensate particle number $n_{cl}=|\Phi_l|^2$ within the Gutzwiller approximation. 
	Using Eq.~(\ref{eq:EOM_GutzwillerCoefficient}) and Eq.~(\ref{eq:OrderParameter}), we obtain
	\begin{align}
	&\frac{dn_{cl}}{dt}
	=I_{Jl}+I_{Ul},
	\label{eq_CC}
	\end{align}
	where 
	\begin{align}
	&I_{Jl}
	=
	\frac{J}{i\hbar}\sum_{\braket{m}_l}\left(\Phi_l\Phi_m^\ast-\Phi_l^\ast\Phi_m\right) ,\\
	&I_{Ul}
	=
 	\frac{U}{i\hbar}\sum_{n}n\sqrt{n+1}\left(f_{n}^{(l)\ast}f_{n+1}^{(l)}\Phi_l^\ast-f_{n}^{(l)}f_{n+1}^{(l)\ast}\Phi_l\right) .
	\end{align}
	The first term in Eq.~\eqref{eq_CC} represents the particle current arising from hopping, which conserves the total condensate particle number. 
	The second term results from the conversion from the condensate to the non-condensate and vice versa.
	In the weakly interacting, deep superfluid state, 
	the Gutzwiller coefficient $f_n^{(l)}$ has a weak $n$ dependence,
	and thus the contribution of the second term becomes very small.
	In this limit, the first term reduces to the usual expression for the supercurrent for the discrete GP equation.
	In contrast, near the Mott lobe, the contribution of the second term becomes relatively large, as shown in Fig.~\ref{figure_0_5}.
	In this regime, the conversion between the condensate and non-condensate components plays a crucial role in the condensate dynamics.

	It is important to note that the r.h.s of Eq.~\eqref{eq_CC} vanishes when $\Phi_l=0$, and thus it should not be able to excite the condensate dynamics in the Mott-insulator region within the Gutzwiller approximation.
	Nevertheless, we observe permeation of the condensate due to the soliton propagation.
	This is because in the Mott-insulator region of the MI--SF pseudo-heterojunction system, the local condensate number does not completely vanish, but there remains a small amount of the condensate.

\section{Out-of-phase Soliton injection to MI shell } 
	\begin{figure*}[tb]
	\centering
		\includegraphics[scale=0.117]{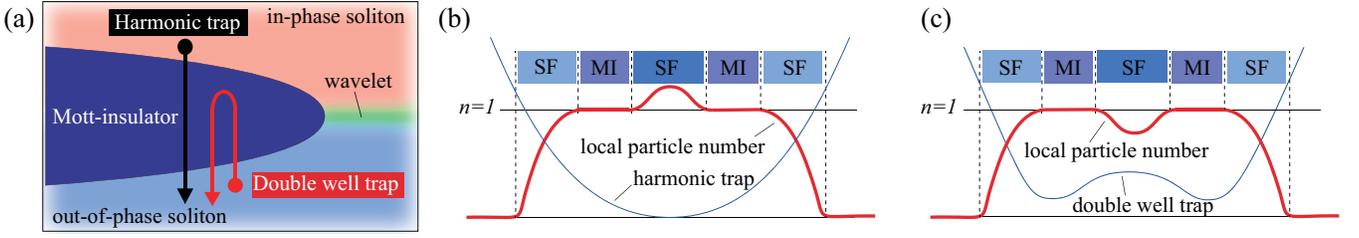}
		\caption{
		Difference between the harmonic trap system and the double well trap system. 
		(a) Schematics of the change of the effective chemical potential in the dynamical phase diagram close to the Mott lobe. 
		The starting points of the two arrows represent the center of the trap. 
		Structures of the local particle number for the harmonic trap system (b) and double well trap system (c). 
		The red and blue lines represent the configurations of the local particle number and external potential, respectively. 
		}
		\label{figure_0_4}
	\end{figure*}

	Figure~\ref{figure_0_4} schematically displays (a) how the chemical potential changes effectively in the dynamical phase diagram and (b) the configurations of the local particle number in the presence of only a harmonic trap. 
	In the previous section, where a harmonic trap potential is applied, the effective chemical potential decreases monotonically as a function of the distance from the center of the trap (represented by the black arrow in (a) of Fig.~\ref{figure_0_4}).
	As a result, we cannot inject the various types of solitons into the outer MI and SF shells, where all the excitation turns into an in-phase soliton before reaching the surface of the MI shell, and hence the in-phase soliton is always injected.
	In the harmonic trap case, the local particle number  in the SF core is larger  than that in the SF shell, and the MI shell has a commensurate filling (see (b) in Fig.~\ref{figure_0_4}).

	An out-of-phase soliton can be injected into the outer MI shell by inducing a non-monotonic external potential, such as an external potential peaking at the center of the harmonic trap, where the effective chemical potential behaves as the red arrow in (a) of Fig.~\ref{figure_0_4}. 
In this section, we apply an additional Gaussian potential 
		\begin{equation}
		V_G=Ae^{-B(l-l_0)^2},
		\end{equation}
		where  we used the coefficients $A=0.3$, $B=0.01$, $J/U=0.075$, and $\mu/U=0.5$ in the following calculations.
		The width of the SF core is narrowed due to the presence of the repulsive potential that is applied to the center of the harmonic trap.
	In order to broaden the width of the SF core, we set a value $V_0/U=2.5\times 10^{-4}$, which realizes a gradual curvature of the harmonic trap.
	By applying the additional Gaussian potential, the local particle number in the SF core and SF shell is less than the commensurate filling $n=1$ (see Fig.~\ref{figure_0_4}(c)). 
	 With these parameters, we numerically verified that an out-of-phase soliton is created at the center of the trap  and the structure of the out-of-phase soliton is maintained until the soliton collides with the surface of the MI shell ($n=1$). 
	
\begin{figure}[tb]
	\centering
		\includegraphics[scale=0.7]{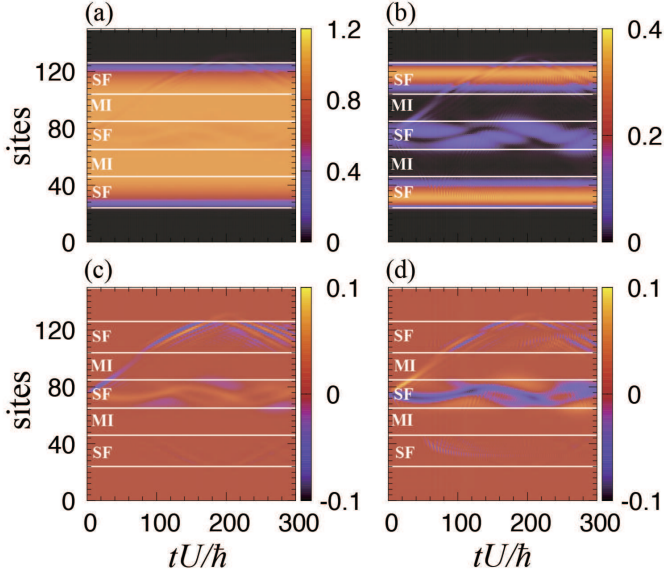}
		\caption{
		The same plots as in Fig.~\ref{figure_1_1} but for the condition where an additional Gaussian potential is introduced. 
		We used $J/U=0.075$ and $\mu/U=0.5$. 
		}
		\label{figure_2_1}	
	\end{figure}

	In Fig.~\ref{figure_2_1} (a) and (b), we plot the local particle number and condensate particle number as a function of the site and time. 
	We can clearly observe that the out-of-phase soliton propagates to the outer MI shell while keeping its characteristic features (Fig.~\ref{figure_2_1} (c) and (d)).
	In this situation, the condensate permeates the MI regions and the excitation then also permeates the exterior SF regions. 
	This contrasts with the case of only the harmonic trap, where the excitations are reflected at the surface of the exterior SF shell. 
	 
	An additional repulsive Gaussian potential with the harmonic trap provides a result where the excitation permeates into the outer SF shell. 
	This contrasts with the case with only the harmonic trap, where the excitation cannot permeate into the outer SF shell. 
	This difference can be attributed to a match between the effective chemical potential at the SF core and that of the outer SF shell. 
	We confirm this fact by using a system with a simple model of a staircase potential, where the MI shell and SF shell are displayed alternatively and the chemical potential is constant within each shell. 
	We find that not only the out-of-phase soliton but also the other types of solitons can be injected into the outer SF shell in the case where the chemical potential or local particle number in two SF shells match. 
	We therefore conclude that an impedance match between the effective chemical potential or local particle number between the SF core and outer SF shell is a crucial factor for excitation injection into the outer SF shell, which is very similar to impedance matching between two BECs in the context of the transmission of the Bogoliubov excitation~\cite{Watabe2008}. 
	
	  Finally, we briefly report on sound propagation, i.e., the propagation of the density fluctuation created by applying the local single-shot pulse potential at the SF core. 
	   Immediately after the pulse is imposed, two dips in the local particle number propagate in opposite directions with the same speed. 
	    We find that the properties of the excitation injection are the same as the soliton cases.
	    In the harmonic trap case, the excitation cannot be injected into the outer SF shell, 
	    and is repelled at the surface of the SF shell after the excitation permeates through the MI shell. 
	    On the other hand, the excitation can be injected into the outer SF shell when an additional external potential peaking at the center of the harmonic trap is introduced. 
	    This is analogous to impedance matching between two BECs in the context of the transmission of the Bogoliubov excitation~\cite{Watabe2008}. 
	    These results indicate that if we experimentally study the impedance matching of excitations between the SF core and outer SF shell, we can choose whichever of the two setups is more convenient: the phase-imprinting method or local single-shot pulse potential.

\section{conclusion}	

We have investigated the properties of phase-imprinted solitons of ultracold bosons in an optical lattice using the time-dependent Gutzwiller approximation.
Three types of phase-imprinted solitons in the Bose--Hubbard model have been reported in an earlier paper~\cite{Krutitsky2010}, namely, the in-phase soliton, out-of-phase soliton, and wavelet. 
In this paper, we determined the dynamical phase diagram 
and found a new type of soliton in addition to the three solitons reported in Ref.~\cite{Krutitsky2010}. 
We call this new soliton the hybrid soliton. 
It has a single peak and single dip in the local particle number but two dips in the local condensate particle number. 

Although four types of phase-imprinted solitons can be created at the center of the harmonic trap in an optical lattice system, 
the shape of all the solitons deforms into an in-phase soliton before reaching the outer MI shell, 
and an in-phase soliton is always injected to the outer MI shell. 
This is because the effective chemical potential monotonically decreases further from the center of the harmonic trap, 
and the upper side of the MI lobe always faces the in-phase soliton region in the dynamical phase diagram.  
After the solitons collide with the surface of the MI shell, the condensate permeates into the outer MI shell. 
However, the excitations cannot penetrate into the outer SF region and are reflected by its surface. 
In the presence of an additional repulsive potential peaking at the center of the harmonic trap, 
the effective chemical potential varies non-monotonically from the center of the trap 
and we can inject excitations into the outer SF shell. 
These properties can be observed for phase-imprinted solitons as well as for the sound wave induced by the local single-shot pulse potential, which can be related to impedance matching of excitations in BECs in terms of the effective chemical potential or local particle number in the shell. 
We hope that the present study on the heterojunction of SF--MI--SF states will be useful for the development of atomtronics analogous to the Josephson junction in superconductivity, which plays a crucial role for the development of superconducting qubits.

\begin{acknowledgments}
SW was supported by JSPS KAKENHI Grant No. JP18K03499. 
\end{acknowledgments} 

\appendix	
\section{}
In this appendix, we briefly show how the $d$-dimensional Bose--Hubbard model reduces to the one-dimensional Gutzwiller equation.
We start with the $d$-dimensional Bose--Hubbard Hamiltonian
	\begin{align}
	\hat{H}&=-\sum_\mathbf{i}\sum_{\alpha=1}^d J_\alpha(\hat{b}^\dagger_\mathbf{i}\hat{b}_{\mathbf{i}+\mathbf{e}_\alpha}+\mathrm{H.c.})+\sum_\mathbf{i}(\epsilon_\mathbf{i}-\mu)\hat{n}_\mathbf{i} \nonumber \\
	&\hspace{1cm}+\frac{U}{2}\sum_\mathbf{i}\hat{n}_\mathbf{i}(\hat{n}_\mathbf{i}-1), 
	\end{align}
where $\mathbf{e}_\alpha$ is a unit vector in the direction $\alpha$ and $J_\alpha$ is the anisotropic hopping coefficient that depends on $\alpha$. 
In general, the Gutzwiller ansatz  is written as
	\begin{align}
	\ket{\Phi}=\prod_\mathbf{i}\sum_n f_n^{(\mathbf{i})} .
	\end{align}
The equation of motion for the Gutzwiller coefficient directly derived from the 3D Bose--Hubbard Hamiltonian is given by
	\begin{align}
	i\hbar \frac{df^{(\mathbf{i})}_n}{dt}
	&=-(\bar{\Phi}_\mathbf{i}\sqrt{n}f_{n-1}^{(\mathbf{i})}+\bar{\Phi}_\mathbf{i}^{\ast}\sqrt{n+1}f_{n+1}^{(\mathbf{i})}) \nonumber\\
	&\hspace{1cm}+\left[\frac{U}{2}n(n-1)+(\epsilon_\mathbf{i}-\mu)n\right]f_n^{(\mathbf{i})},
	\end{align}
where $\bar{\Phi}_\mathbf{i}=\sum_{\alpha=1}^dJ_\alpha(\Phi_{\mathbf{i}+\mathbf{e}_\alpha}+\Phi_{\mathbf{i}-\mathbf{e}_\alpha})$.
Here we assume that the external trap potential is applied in one direction $\alpha=1$ and thus the on-site energy $\epsilon_\mathbf{i}$ depends only on $i_1$.
We also assume that the Gutzwiller coefficient $f^{(\mathbf{i})}_n$ depends only on one spatial dimension $\alpha=1$
and hence the order parameter depends only on the direction $\alpha=1$.
Denoting $i_1=l$, $\bar{\Phi}_\mathbf{i}^\alpha$ is rewritten as
	\begin{align}
	\bar{\Phi}_\mathbf{i}=
		\begin{cases}
		J_1(\Phi_{l+1}+\Phi_{l-1}) & \alpha=1,\\
		2J_\alpha\Phi_{l} & \alpha\neq 1.
		\end{cases}
	\end{align}
We thus obtain
	\begin{align}
	i\hbar\frac{df^{(l)}_n}{dt}
	&=-J\left[(\Phi_{l+1}+\Phi_{l-1})\sqrt{n}f_{n-1}^{(l)}\right. \nonumber\\
	&\left.+(\Phi_{l+1}^\ast+\Phi_{l-1}^\ast)\sqrt{n+1}f_{n+1}^{(l)}\right] \nonumber\\
	&-2J^\prime\left(\Phi_l\sqrt{n}f_{n-1}^{(l)}+\Phi_l^\ast\sqrt{n+1}f_{n+1}^{(l)}\right)\nonumber\\
	&\hspace{1cm}+\left[\frac{U}{2}n(n-1)+(\epsilon_l-\mu)n\right]f_n^{(l)},
	\label{eq_appendix}
	\end{align}
where we have denoted $J=J_1$ and defined $J^\prime\equiv\sum_{\alpha\neq 1}^d J_\alpha$.
The one-dimensional Gutzwiller equation~(\ref{eq:EOM_GutzwillerCoefficient}) corresponds to the case where the second term of Eq.~(\ref{eq_appendix}) is neglected.
It is well known that for the uniform BH model, the role of the second term in Eq.~(\ref{eq_appendix}) in the equilibrium phase diagram is to replace the hopping coefficient as $J \rightarrow \sum_\alpha J_\alpha$.
On the other hand, this term will not have a significant effect on the dynamics accompanying the spatial variation in the direction $\alpha=1$.
For instance, this term does not create any distributions in the equation of motion (see Eq.~(11)) for the local condensate particle number $n_{cl}$.
We thus expect that our analysis using the one-dimensional Gutzwiller equation~(\ref{eq:EOM_GutzwillerCoefficient}) well captures the essential physics related to soliton propagation for realistic three-dimensional systems.
In fact, our results for the uniform Bose--Hubbard model are consistent with the results obtained in Ref.~\cite{Krutitsky2010}, which used Eq.~(\ref{eq_appendix}).
	
\bibliographystyle{apsrev4-1}
\bibliography{dark_soliton_BH}

\end{document}